\documentclass[prl,aps,superscriptaddress,twocolumn,notitlepage,showpacs]{revtex4-1}
\usepackage{latexsym}
\usepackage{amsmath}
\usepackage{amssymb}
\usepackage{graphicx}
\usepackage{caption}
\usepackage{subfigure}
\usepackage{float}
\usepackage{mathrsfs}
\usepackage{color}
\usepackage{txfonts}
\usepackage[justification=centering,
            format=plain]{caption}

\renewcommand{\raggedright}{\leftskip=0pt \rightskip=0pt plus 0cm}

\begin{document}

\title{Quantum entanglement between two magnon modes via Kerr nonlinearity}

\author{Zhedong Zhang}
\email{zhedong.zhang@tamu.edu}
\affiliation{Institute for Quantum Science and Engineering, Texas A$\&$M University, College Station, TX 77843, USA}

\author{Marlan O. Scully}
\affiliation{Institute for Quantum Science and Engineering, Texas A$\&$M University, College Station, TX 77843, USA}
\affiliation{Quantum Optics Laboratory, Baylor Research and Innovation Collaborative, Waco, TX 76704, USA}
\affiliation{Department of Mechanical and Aerospace Engineering, Princeton University, Princeton, NJ 08544, USA}

\author{Girish S. Agarwal}
\email{girish.agarwal@tamu.edu}
\affiliation{Institute for Quantum Science and Engineering, Texas A$\&$M University, College Station, TX 77843, USA}
\affiliation{Department of Biological and Agricultural Engineering, Department of Physics and Astronomy, Texas A$\&$M University, College Station, TX 77843, USA}

\date{\today}

\begin{abstract}
We propose a scheme to entangle two magnon modes via Kerr nonlinear effect when driving the systems far-from-equilibrium. We consider two macroscopic yttrium iron garnets (YIGs) interacting with a single-mode microcavity through the magnetic dipole coupling. The Kittel mode describing the collective excitations of large number of spins are excited through driving cavity with a strong microwave field. We demonstrate how the Kerr nonlineraity creates the entangled quantum states between the two macroscopic ferromagnetic samples, when the microcavity is strongly driven by a blue-detuned microwave field. Such quantum entanglement survives at the steady state. Our work offers new insights and guidance to designate the experiments for observing the entanglement in massive ferromagnetic materials. It can also find broad applications in macroscopic quantum effects and magnetic spintronics.
\end{abstract}

\maketitle

{\it Introduction.--} Recent advance in ferromagnetic materials draw considerable attention in the studies of quantum nature in magnetic systems, as the limitations of electrical circuitry are reached. Thanks to the low loss of the collective excitations of spins known as magnons in magnetic samples, the magnons offer a new paradigm for developing future generation of spintronic devices and quantum engineering \cite{Kajiwara_Nature2010,Cornelissen_NatPhys2015,Zhu_APL2016,An_NatMater2013,Chumak_NatPhys2015,Chumak_NatCommun2014}. The yttrium iron garnet (YIG) with the size of $\sim 100\mu$m as fabricated in recent experiments provides new insights for studying the macroscopic quantum effects, such as entanglement and squeezing that have raised widespread interest in different branches of physics during decade \cite{Collet_NatCommun2016,Ho_PRL2018,Simon_Nat2018,Yuan_PRB2018,Morimae_PRA2005,Korppi_Nature2018}. Quantum entanglement between massive mirror and optical cavity photons has been explored, in both theoretical and experimental aspects \cite{Vitali_PRL2007,Simon_Nature2009,Aspelmeyer_RMP2014,Genes_PRA2008,Verhagen_Nature2012,Palomaki_Sci2013}. Several ideas follow-on suggest the extension of such entangled quantum state towards the magnons in microwave regime, due to their great potential for macroscopic spintronic devices. Much experimental efforts have been devoted to the quantum nature of magnon states, through hybridizing the spin waves with other degrees of freedoms, e.g., superconducting qubits and phonon modes \cite{Julsgaad_Nat2001,Wellstood_Science2003,Quirion_SciAdv2017,Zhang_SciAdv2016}. Compared to atoms and photonics, magnonics holds the potential for implementing quantum states in more massive objects. This can be seen from the $320\mu$m-diam YIG spheres implemented in recent experiments \cite{Zhang_NPJ2015}. 

As a powerful platform for investigating the light-matter interaction \cite{Zhang_NPJ2015,Tabuchi_PRL2014,Soykal_PRL2010,Zhang_PRL2014,Bourhill_PRB2016,Tabuchi_Sci2015,Harder_PRL2018,Yao_NatCommun2017}, ferromagnetic materials are taking the advantage of reaching strong and ultrastrong coupling regimes, along with the fact of their high spin density as well as low dissipation rate. The strong coupling results in the cavity magnon-polariton, serving as a potential candidate for implementing quantum information transducers and memories \cite{Yao_NatCommun2017,Zhang_NatCommun2015}. 
To achieve the quantum regime in magnon polaritons, the macroscopic quantum effects are essentially worthy of being explored. The most recent work using driven-dissipation theory suggest the magnon-photon-phonon entanglement and also the squeezing of magnon modes in which both the entanglement and squeezing are essentially transferred into the mechanical mode \cite{Li_PRL2018,Li_PRA2019,Li_magnon2019}. 
From a theoretical view-point, this macroscopic quantum nature of magnon modes stems from the nonlinearity that can be enhanced by driving the systems far-from-equilibrium. Two prominent schemes are responsible for introducing such nonlinearity: the magnetostrictive interaction and the Kerr effect, where the latter results from the magnetocrystalline anisotropy. Apart from the magnon-phonon interaction, Kerr nonlinearity plays a significant role in magnon spintronics \cite{Chumak_NatPhys2015}. Recent experiments in YIG spheres demonstrated the multistability and photon-mediated control of spin current, due to the Kerr effect \cite{You_PRL2018,Hu_PRL2017}.

In this Letter, we propose a scheme of entangling magnon modes in two massive YIG spheres via the Kerr nonlinearity. The two magnon modes interact with a microcavity through the beam-splitter-like coupling, which cannot produce any entanglement. Nevertheless, activating the Kerr nonlinearity via strong driving results in squeezing-like coupling which may let magnon get entangled with cavity photons. The subsequent entanglement transfer between photons and the other magnon mode will lead to the entanglement between magnon modes. The condition for optimizing the magnon-magnon entanglement is found and is confirmed by our numerical calculations. By taking into account the experimentally feasible parameters, we show the considerable magnon-magnon entanglement can be created. Such entanglement is also shown to be robust against cavity leakage. Our work offers new insight and perspective for studying the quantum effects in complex molecules. These have been manifested by the excited-state dynamics in dye molecules and even bacterias implying the entangled quantum states when interacting with microcavities \cite{Sarovar_NatPhys2010,Zhang_SR2016,Coles_NatMater2014,Zhang_JCP2018,Mukamel_JPCL2016}.

\begin{figure}
 \captionsetup{justification=raggedright,singlelinecheck=false}
 \centering
   \includegraphics[scale=0.22]{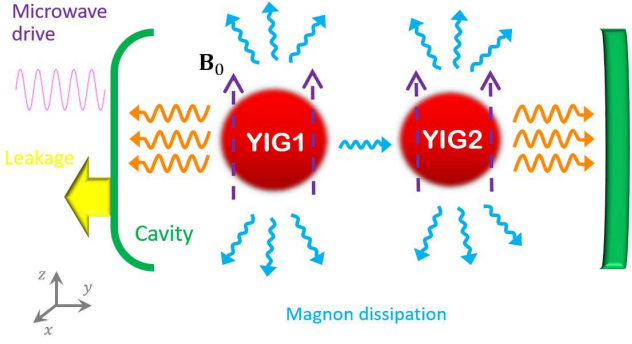}
\caption{Schematic of cavity magnons. Two YIG spheres are interacting with the basic mode of microcavity in which the right mirror is made of high-reflection material so that photons leak from the left side. The static magnetic field for producing Kittel mode is along $z$-axis whereas the microwave driving and magnetic field inside cavity are along $x$-axis.}
\label{sch}
\end{figure}

{\it Model and equation of motion.--} We consider a hybrid magnon-cavity system consisting of two bulk ferromagnetic materials and one microwave cavity mode. The ferromagnetic sample contains dispersive spin waves, in which only the spatially uniform mode (Kittel mode \cite{Kittel_PR1948}) is assumed to strongly interact with cavity photons. The full Hamiltonian of this cavity magnonics system reads \cite{Blundell_book2001}
\begin{equation}
\begin{split}
H = & -\int M_z B_0\text{d}\textbf{r} - \frac{\mu_0}{2}\int M_z H_{\text{an}}\text{d}\textbf{r}\\[0.2cm]
& + \frac{1}{2}\int\left(\varepsilon_0\textbf{E}^2+\frac{\textbf{B}^2}{\mu_0}\right)\text{d}\textbf{r} - \int \textbf{M}\cdot\textbf{B}\text{d}\textbf{r}
\end{split}
\label{H}
\end{equation}
where $\textbf{B}_0=B_0\textbf{e}_z$ is the applied static magnetic field and $\textbf{M}=\gamma\textbf{S}/V_m$ with $\gamma=e/m_e$ denoting the  gyromagnetic ratio. $\textbf{S}$ stands for the collective spin operator and $V_m$ is volume of ferromagnetic material. $\textbf{H}_{\text{an}}$  is the anisotropic field due to the magnetocrystalline anisotropy and has $z$ component only owing to the crystallographic axis being aligned along the applied static magnetic field. Thereby the anisotropic field is given by $H_{\text{an}}=-2K_{\text{an}}M_z/M^2$ where $K_{\text{an}}$ and $M$ denote the dominant 1st anisotropy constant and the saturation magnetization, respectively. One can recast the Hamiltonian in Eq.(\ref{H}) into
\begin{equation}
\begin{split}
H = -\gamma & \sum_{j=1}^2 B_{j,0} S_{j,z} + \gamma^2\sum_{j=1}^2\frac{\mu_0 K_{\text{an}}^{(j)}}{M_j^2 V_{j,m}}S_{j,z}^2\\
& + \hbar\omega_c a^{\dagger}a - \gamma \sum_{j=1}^2 S_{j,x} B_{j,x}
\end{split}
\label{H2}
\end{equation}
by assuming the magnetic field inside cavity is along $x$-axis. The Holstein-Primakoff transform yields to $S_{i,z}=S_i-m_i^{\dagger}m_i,\ S_{i,+}=(2S_i-m_i^{\dagger}m_i)^{1/2}m_i,\ S_{i,-}=m_i^{\dagger}(2S_i-m_i^{\dagger}m_i)^{1/2}$ where $S_{i,\pm}\equiv S_{i,x}\pm i S_{i,y}$ and $m_i$ represents the bosonic annihilation operator \cite{Madelung_book1978}. For the yttrium iron garnets (YIGs) with diameter $\text{d}=40\mu$m, the density of ferrum ion $\text{Fe}^{3+}$ is $\rho=4.22\times 10^{27}$m$^{-3}$, which leads to the total spin $S=\frac{5}{2}\rho V_m=7.07\times 10^{14}$. This is often much larger than the number of magnons, so that we can safely approximate $S_{j,+}\simeq \sqrt{2S_j}m_j,\ S_{j,-}\simeq\sqrt{2S_j}m_j^{\dagger}$. In the presence of external microwave driving field, the effective Hamiltonian of hybrid magnon-cavity system is of the form
\begin{equation}
\begin{split}
H_{\text{eff}} = \hbar & \omega_c a^{\dagger}a + \hbar\sum_{j=1}^2 \Big[\omega_j m_j^{\dagger}m_j + g_j(m_j^{\dagger}a+m_j a^{\dagger})\\[0.15cm]
& + \Delta_j m_j^{\dagger}m_j m_j^{\dagger}m_j\Big] + i\hbar\Omega(a^{\dagger}e^{-i\omega_{\text{d}} t}-a e^{i\omega_{\text{d}} t})
\end{split}
\label{Heff}
\end{equation}
where the rotating-wave approximation was employed and cavity frequency is denoted by $\omega_c$. The frequency of Kittel mode is $\omega_j=\gamma B_{j,0}$ with $\gamma/2\pi=28$GHz/T. $g_j$ gives the magnon-cavity coupling and $\Delta_j=\mu_0 K_{\text{an}}^{(j)}\gamma^2/M_j^2 V_{j,m}$ gives the Kerr nonlinearity, resulting from the on-site magnon-magnon scattering. The Rabi frequency $\Omega=\sqrt{2 P_{\text{d}}\gamma_c/\hbar\omega_{\text{d}}}$ in last term quantifies the strength of the field inside microcavity driven by the microwave magnetic field, where $P_{\text{d}}$ and $\omega_{\text{d}}$ represent the power and frequency of the microwave field, respectively. $\gamma_c$ is the cavity leaking rate. In the rotating frame of microwave field, the dynamics of hyrid cavity-magnon system is governed by the quantum Langevin equations (QLEs)
\begin{equation}
\begin{split}
& \dot{m}_s = -(i\delta_s + \gamma_s)m_s - 2i\Delta_s m_s^{\dagger}m_s m_s -ig_s a + \sqrt{2\gamma_s}m_s^{\text{in}}(t)\\[0.15cm]
& \dot{a} = -(i\delta_c + \gamma_c)a - i\sum_{j=1}^2 g_j m_j + \Omega + \sqrt{2\gamma_c}a^{\text{in}}(t)
\end{split}
\label{qle}
\end{equation}
where $\gamma_s$ quantifies the magnon dissipation. $\delta_s=\omega_s-\omega_{\text{d}},\ \delta_c=\omega_c-\omega_{\text{d}}$. 
$m_s^{\text{in}}(t)$ and $a^{\text{in}}(t)$ are the input noise operators having zero mean and white noise: $\langle m_s^{\text{in},\dagger}(t)m_s^{\text{in}}(t')\rangle=\bar{n}_s\delta(t-t'),\ \langle m_s^{\text{in}}(t)m_s^{\text{in},\dagger}(t')\rangle=(\bar{n}_s+1)\delta(t-t');\ \langle a^{\text{in},\dagger}(t)a^{\text{in}}(t')\rangle=0,\ \langle a^{\text{in}}(t)a^{\text{in},\dagger}(t')\rangle=\delta(t-t')$ where $\bar{n}_s=[\text{exp}(\hbar\omega_s/k_B T)-1]^{-1}$ denotes the Planck factor of the $s$-th magnon mode.

\begin{figure*}[t]
 \captionsetup{justification=raggedright,singlelinecheck=false}
 \centering
   \includegraphics[scale=0.32]{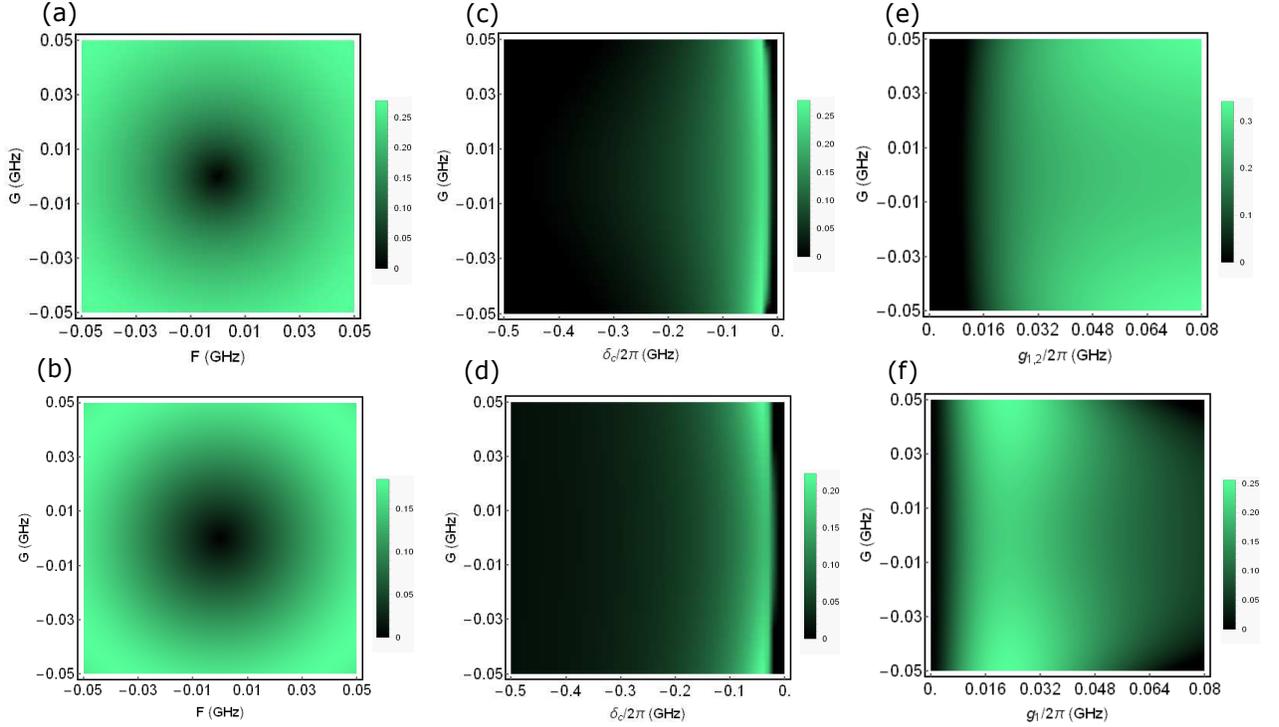}
\caption{2D plots for (top) magnon-magnon entanglement $E_{m_1m_2}$ and (bottom) magnon-cavity entanglement $E_{m_1 a}$ when turning off the coupling between cavity and the 2nd sphere ($g_2=0$). (a) $g_{1,2}/2\pi=41$MHz, $\delta_c/2\pi=-0.03$GHz; (b) $g_1/2\pi=41$MHz, $g_2=0$, $\delta_c/2\pi=-0.03$GHz; (c) $F_{1,2}=-0.048$GHz, $g_{1,2}/2\pi=41$MHz; (d) $F_{1,2}=-0.048$GHz, $g_1/2\pi=41$MHz, $g_2=0$ and (e,f) $F_{1,2}=-0.048$GHz, $\delta_c/2\pi=-0.03$GHz. Other parameters are $\omega_{1,2}/2\pi=10$GHz, $\delta_{1,2}/2\pi=-1$MHz, $\gamma_{1,2}/2\pi=8.8$MHz, $\gamma_c/2\pi=1.9$MHz and $T=10$mK.}
\label{E2D}
\end{figure*}

Since the microcavity is under strong driving by the microwave field, the beam-splitter-like coupling between magnons and cavity leads to the large amplitudes of both magnon and cavity modes, namely, $|\langle m_s\rangle|,\ |\langle a\rangle|\gg 1$. In this case, one can  safely introduce the expansion $m_s=\langle m_s\rangle+\delta m_s,\ a=\langle a\rangle+\delta a$ in the vicinity of steady state, by neglecting the higher-order fluctuations of the operators. We thereby obtain the linearized QLEs for the quadratures $\delta X_s,\delta Y_s,\delta X,\delta Y$ defined as $\delta X_1=(\delta m_1+\delta m_1^{\dagger})/\sqrt{2},\ \delta Y_1=(\delta m_1-\delta m_1^{\dagger})/i\sqrt{2},\ \delta X_2=(\delta m_2+\delta m_2^{\dagger})/\sqrt{2},\ \delta Y_2=(\delta m_2-\delta m_2^{\dagger})/i\sqrt{2},\ \delta X=(\delta a+\delta a^{\dagger})/\sqrt{2},\ \delta Y=(\delta a-\delta a^{\dagger})/i\sqrt{2}$
\begin{equation}
\begin{split}
\dot{\sigma}(t) = A\sigma(t) + f(t)
\end{split}
\label{u}
\end{equation}
where $\sigma(t)=[\delta X_1(t), \delta Y_1(t), \delta X_2(t), \delta Y_2(t),\delta X(t), \delta Y(t)]^{\text{T}}$ and $f(t){=}[\sqrt{2\gamma_1}X_1^{\text{in}}(t),\! \sqrt{2\gamma_1}Y_1^{\text{in}}(t),\! \sqrt{2\gamma_2}X_2^{\text{in}}(t),\! \sqrt{2\gamma_2}Y_2^{\text{in}}(t),\! \sqrt{2\gamma_c}X^{\text{in}}(t),\! \\ \sqrt{2\gamma_1}Y^{\text{in}}(t)]^{\text{T}}$ are the vectors for quantum fluctuations and noise, respectively. The drift matrix reads
\begin{equation}
\begin{split}
A = \begin{pmatrix}
     F_1-\gamma_1 & \tilde{\delta}_1-G_1 & 0 & 0 & 0 & g_1\\[0.15cm]
		 -\tilde{\delta}_1-G_1 & -F_1-\gamma_1 & 0 & 0 & -g_1 & 0\\[0.15cm]
		 0 & 0 & F_2-\gamma_2 & \tilde{\delta}_2-G_2 & 0 & g_2\\[0.15cm]
		 0 & 0 & -\tilde{\delta}_2-G_2 & -F_2-\gamma_2 & -g_2 & 0\\[0.15cm]
		 0 & g_1 & 0 & g_2 & -\gamma_c & \delta_c\\[0.15cm]
		 -g_1 & 0 & -g_2 & 0 & -\delta_c & -\gamma_c
    \end{pmatrix}
\end{split}
\label{A}
\end{equation}
with magnetocrystalline anisotropy quantified by $G_s=2\Delta_s\text{Re}\langle m_s\rangle^2,\ F_s=2\Delta_s\text{Im}\langle m_s\rangle^2$ and the effective detuning of magnons $\tilde{\delta}_s=\delta_s+2\sqrt{G_s^2+F_s^2}=\delta_s+4\Delta_s|\langle m_s\rangle|^2$, which includes the frequency shift caused by Kerr nonlinearity. The mean $\langle m_{1,2}\rangle$ are given by
\begin{equation}
\begin{split}
& \langle m_1\rangle = \frac{ig_1\Omega}{(\tilde{\delta}_1-i\gamma_1)(\delta_c-i\gamma_c)-g_1^2-\frac{g_2^2(\tilde{\delta}_1-i\gamma_1)}{\tilde{\delta}_2-i\gamma_2}},\\[0.15cm]
& \text{and}\quad (1\leftrightarrow 2)
\end{split}
\label{m}
\end{equation}
Before the study of entanglement, it is essential to elucidate the mechanism for optimizing the entanglement via Kerr nonlinearity. To this end, we proceed via the effective Hamiltonian for quantum fluctuations
\begin{equation}
\begin{split}
H_{\text{qf}} = \hbar\sum_{s=1}^2 & \Big[\tilde{\delta}_s\delta m_s^{\dagger}\delta m_s + \tilde{\Delta}_s\delta m_s^{\dagger}\delta m_s^{\dagger} + \tilde{\Delta}_s^*\delta m_s\delta m_s\\[0.15cm]
& + g_s\left(\delta m_s^{\dagger}\delta a + \delta m_s \delta a^{\dagger}\right)\Big] + \hbar\delta_c\delta a^{\dagger}\delta a
\end{split}
\label{Heff}
\end{equation}
where $\tilde{\Delta}_s=(G_s+iF_s)/2$. The quadratic terms $\delta m_s^{\dagger}\delta m_s^{\dagger},\ \delta m_s\delta m_s$ imply the effective magnon-magnon interaction induced by the magnetocrystalline anisotropy, which may be significantly enhanced by strong driving. This, in fact, is responsible for the entanglement. To make it elaborate, let us introduce the Bogoliubov transformation \cite{Lifshitz_book1980,Fetter_book2003} $\delta\beta_s=u_s\delta m_s-v_s^*\delta m_s^{\dagger},\ \delta\beta_s^{\dagger}=-v_s\delta m_s+u_s^*\delta m_s^{\dagger}$ where $u_s=\sqrt{\frac{1}{2}\left(\frac{\tilde{\delta_s}}{\varepsilon_s}+1\right)},\ v_s e^{i\alpha}=-\sqrt{\frac{1}{2}\left(\frac{\tilde{\delta_s}}{\varepsilon_s}-1\right)},\ \alpha=\text{arctan}(F_s/G_s)$ and $\varepsilon_s=\left(\tilde{\delta}_s^2-4|\tilde{\Delta}_s|^2\right)^{1/2}$. Inserting these into Eq.(\ref{Heff}) we find
\begin{equation}
\begin{split}
H_{\text{qf}} = \hbar & \sum_{s=1}^2 \Big[\varepsilon_s\delta \beta_s^{\dagger}\delta \beta_s + g_s\Big((v_s\delta\beta_s+u_s\delta\beta_s^{\dagger})\delta a\\[0.15cm]
& \quad + (u_s^*\delta\beta_s+v_s^*\delta\beta_s^{\dagger})\delta a^{\dagger}\Big)\Big] + \hbar\delta_c\delta a^{\dagger}\delta a
\end{split}
\label{Heffb}
\end{equation}
which shows $\varepsilon_s\simeq -\delta_c$ is optimal for the entanglement, due to the magnon-photon squeezing term $g_s(v_s\delta\beta_s\delta a+v_s^*\delta\beta_s^{\dagger}\delta a^{\dagger})$. This will be confirmed by the latter numerical results when taking into account of experimental parameters.

{\it Entanglement between magnon modes.--} Since we are using the linearized quantum Langevin equations, the Gaussian nature of the input states will be preserved during the time evolution of systems. The quantum fluctuations are thus the continuous three-mode Gaussian state, which is completely characterized by an $6\times 6$ covariance matrix (CM) defined as $C_{ij}(t,t')=\frac{1}{2}\langle \sigma_i(t)\sigma_j(t')+\sigma_j(t')\sigma_i(t)\rangle;\ (i,j=1,2,\cdots,6)$ where the average is taken over the system and bath degrees of freedoms. Suppose the drift matrix $A$ is negatively defined, the solution to Eq.(\ref{u}) is $\sigma(t)=M(t)\sigma(0)+\int_0^t M(s)f(t-s)\text{d}s$ where $M(t)=\text{exp}(At)$. This enables us to find the equation which CM obeys
\begin{equation}
\begin{split}
\dot{C}(t+\tau,t) = A C(t+\tau,t) + C(t+\tau,t)A^{\text{T}} + e^{A\tau}D
\end{split}
\label{CM}
\end{equation}
for $\tau\ge 0$. Thus the stationary CM can be straightforwardly obtained by letting $\tau=0,\ t\rightarrow\infty$ in Eq.(\ref{CM}) that yields to the Lyapunov equation
\begin{equation}
\begin{split}
A C_{\infty} + C_{\infty} A^{\text{T}} = -D
\end{split}
\label{CMss}
\end{equation}
where the diffusion matrix is $D=\text{diag}[\gamma_1(2\bar{n}_1+1),\gamma_1(2\bar{n}_1+1),\gamma_2(2\bar{n}_2+1),\gamma_2(2\bar{n}_2+1),\gamma_c,\gamma_c]$ defined through $\langle f_i(t)f_j(t')+f_j(t')f_i(t)\rangle=2D_{ij}\delta(t-t')$. We adopt the logarithmic negativity $E_N$ to quantify the magnon-magnon and magnon-photon entanglements by comupting the $4\times 4$ CM related to the two magnon modes. This can be achieved by defining $E_N=\text{max}[0,-\text{ln}2v_-]$ where $v_-=\text{min}|\text{eig}\oplus_{j=1}^2 (-\sigma_y) P_{12}C_{\infty} P_{12}|$ and $\sigma_y$ is the Pauli matrix \cite{Vidal_PRA2002,Simon_PRL2000}. The matrix $P_{12}=\sigma_z\oplus 1$ realizes the partial transposition at the level of CM. In what follows, we will work in the monostable scheme of magnons. 
Furthermore, we will focus on the case of two identical magnons having $G_{1,2}=G,\ F_{1,2}=F,\ \tilde{\delta}_{1,2}=\tilde{\delta},\ \Delta_{1,2}=\Delta,\ g_{1,2}=g$.

Fig.\ref{E2D} shows the magnon-magnon entanglement versus some key parameters of the system. Here we have taken into account the experimentally feasible parameters \cite{You_PRL2018}: $\omega_{1,2}/2\pi=10$GHz, $\delta_{1,2}/2\pi=-1$MHz, $\gamma_{1,2}/2\pi=8.8$MHz and $\gamma_c/2\pi=1.9$MHz for the YIG bulk at low temperature $T=10$mK. First of all we observe from Fig.\ref{E2D}(a,b) that the Kerr nonlinearity is responsible for creating the steady-state entanglement between two magnon modes, evident by the fact that the entanglement dies out when $G=F=0$. This results from the dominated beam-splitter-interaction between magnon mode and cavity photons, once $G=F=0$. Thereby no magnon-cavity entanglement can be created, as seen in Fig.\ref{E2D}(b). We take the condition $\varepsilon_s\simeq -\delta_c$ for optimizing the magnon-photon entanglement, as illustrated in Fig.\ref{E2D}(d) where $\varepsilon_s\simeq \sqrt{3(G_s^2+F_s^2)}$. The two-mode squeezing term $g_s(v_s\delta\beta_s\delta a+v_s^*\delta\beta_s^{\dagger}\delta a^{\dagger})$ squeezes the joint state between one magnon mode and cavity photons, which results in the partial entanglement in between. Because the same type of interaction occurs when coupling the other magnon mode with cavity, the two distanced magnon modes are expected to be entangled. This is confirmed in Fig.\ref{E2D}(c) manifesting the optimal magnon-magnon entanglement in the vicinity of $\varepsilon_s\simeq -\delta_c$. The elaborate transfer from magnon-photon entanglement to magnon-magnon entanglement is subsequently evident by Fig.1S in supplementary material (SM) that the considerable reduction of magnon-photon entanglement as the coupling of cavity to another sphere is turned on. Since the biparticle entanglement is originated from the Kerr nonlinearity quantified by $G_{1,2}$ and $F_{1,2}$, there must be the interplay between the couplings $G_s,\ F_s$ and $g_s$ which is depicted in Fig.\ref{E2D}(e,f). In Fig.\ref{E2D}(c) we take $g_{1,2}/2\pi=41$MHz and it implies $\delta_c/2\pi\simeq -0.03$GHz for the optimal entanglement $E_{m_1m_2}$. We then adopt the magnitude of $\delta_c$ for plotting Fig.\ref{E2D}(a,b). Using $\sqrt{G^2+F^2}=2\Delta|\langle m\rangle|^2$ and Eq.(\ref{m}) for the $40\mu$m-diam YIG spheres, the optimal entanglement with $|G|=0.038\text{GHz},\ |F|=0.028$GHz (see Fig.\ref{E2D}(a)) yields to the Rabi frequency $\Omega=1.06\times 10^{15}$Hz, corresponding to the drive power $P_{\text{d}}=314$mW. Indeed, the stronger nonlinearity will create more entanglement between the magnon modes. But we have to ensure the negatively defined matrix $A$ given in Eq.(\ref{A}). Also, the experimental feasibility of ultrastrong drive using microwave field needs the consideration.

\begin{figure}
 \captionsetup{justification=raggedright,singlelinecheck=false}
 \centering
   \includegraphics[scale=0.133]{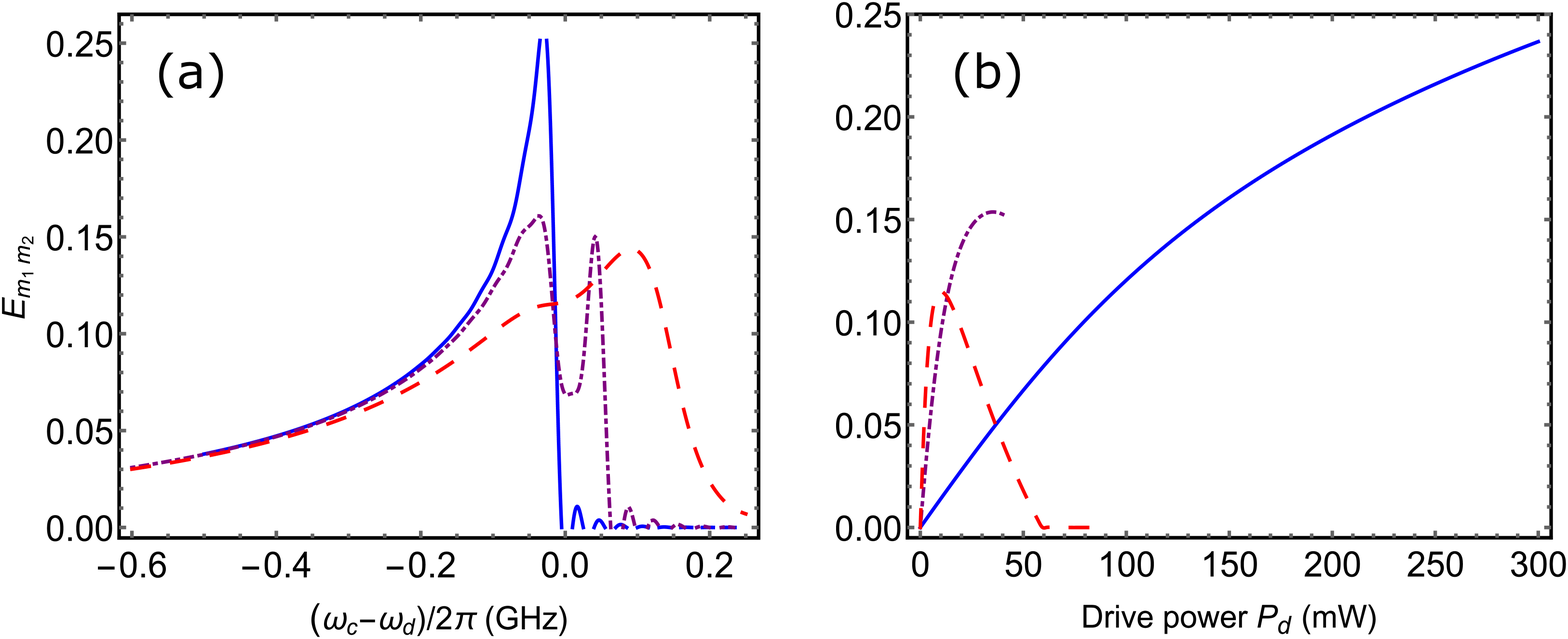}
\caption{Entanglement between two magnon modes varies with (a) cavity detuning and (b) driving power. (a,b) Solid blue, dotdashed purple and dashing red lines are for the cavity leakage $\gamma_c/2\pi=1.9\text{MHz},\ 20\text{MHz}$ and $70\text{MHz}$, respectively; $g_{1,2}/2\pi=41$MHz. (a) Solid blue, dotdashed purple and dashing red lines also correspond to driving power $P_{\text{d}}=393\text{mW},\ 38\text{mW}$ and $11\text{mW}$, respectively; (b) $\delta_c/2\pi=-30$MHz. Other parameters are the same as Fig.\ref{E2D}.}
\label{Elog}
\end{figure}

Since we are working with the strong driving, it is worthy of checking the validity of the results obtained above. The magnon description for magnetic materials is effective only when $\langle m_j^{\dagger}m_j\rangle\ll 2N_js=5N_j$ where $N_j=\rho_j V_{j,m}$ denotes the total number of spins in the bulk material. For the $40\mu$m-diam YIG sphere, $N_j\simeq 1.41\times 10^{14}$ and the drive power $P_{\text{d}}=393$mW results in $|\langle m_j\rangle|\simeq 2.3\times 10^6$, giving $\langle m_j^{\dagger}m_j\rangle\simeq 5.28\times 10^{12}\ll 5N=7.07\times 10^{14}$. Hence the condition $\langle m_j^{\dagger}m_j\rangle\ll 2N_j s$ is fulfilled.

Fig.\ref{Elog} illustrates the entanglement between two magnon modes versus some controllable parameters by considering the $40\mu$m-diam YIG-sphere experiment, where $\omega_{1,2}/2\pi=10$GHz, $\delta_{1,2}/2\pi=-1$MHz, $\Delta_{1,2}/2\pi=1\mu$Hz, $g_{1,2}/2\pi=41$MHz, $\gamma_{1,2}/2\pi=8.8$MHz and $\gamma_c/2\pi=1.9$MHz have been taken according to Ref.\cite{You_PRL2018}. We observe in Fig.\ref{Elog}(a) that for fixed driving power, the magnon-magnon entanglement is quite sensitive to cavity detuning $\delta_c\equiv\omega_c-\omega_{\text{d}}$, reaching its maximum at $\delta_c/2\pi\simeq -0.03$GHz. This is consistent with the condition $\varepsilon_j\simeq -\delta_c$ as clarified for optimizing the entanglement. Fig.\ref{Elog}(b) shows the considerable entanglement when the system is driven far-from-equilibrium. This is reasonable because the strong external driving significantly enhances the Kerr nonlinearity that is responsible for both magnon-cavity squeezing and entanglement, as elucidated in Eq.(\ref{m}) and (\ref{Heff}). Furthermore, Fig.\ref{Elog} shows that the weaker magnon-magnon entanglement is observed when increasing the cavity leakage. By noting the magnitude, we can still obtain some entanglement, even with a low-quality cavity showing weak magnon-cavity coupling where $\gamma_c=8\gamma_{1,2}>g_{1,2}$ denoted by red dashed lines. This regime is crucial for detecting the entanglement used in Refs\cite{Vitali_PRL2007,Palomaki_Sci2013}. in which an additional cavity has a beam-splitter-like interaction with the magnon mode for reading out the magnon states associated with the CM. The transferred entanglement can then be measured through the homodyne detection by sending a weak microwave probe. This approach requires much larger cavity leakage than the magnon dissipation, namely, $\gamma_c\gg\gamma_{1,2}$, so that the magnon states can remain almost unchanged when switching off the laser driving. 

The time-resolved detection of the photons emitting off the cavity axis may offer an alternative scheme for entanglement measurement. The quadrature information of magnon modes can be transferred to the time-gated emitted photons, which can be homodynely detected by interfering with an extra microwave field. This quantum-light-probe scheme may take the advantage of being noninvasive detection for the entanglement measurement.

{\it Conclusion and remarks.--} In conclusion, we have proposed a protocol for entangling the magnon modes in two massive YIG spheres, through the Kerr nonlinearity that originates from the magnetocrystalline anisotropy. We shew that such nonlinearity has to be essentially included, for producing the entanglement. Our work demonstrated the stationary entanglement between two macroscopic YIG spheres driven far-from-equilibrium, within the experimentally feasible parameter regime. The amount of entanglement is quantified by the logarithmic negativity and surprisingly robust against the cavity leakage: entangled quantum state may persist with low-quality cavity giving weak magnon-cavity coupling. This may be helpful to the experimental design for the entanglement measurement.

We should note that our idea for entangling magnon modes may be potentially extensive to other complex systems, such as molecular aggregates and clusters, along with the fact of similar forms of nonlinear couplings $b^{\dagger}b q$ and $\Delta b^{\dagger}bb^{\dagger}b$. With the scaled-up parameters, the long-range entanglement in molecular aggregates would be anticipated, in that the exciton-exciton interaction is of several orders of magnitude higher than the Kerr nonlinearity resulting from the magnetocrystalline anisotropy. For instance, the two-exciton coupling in J-aggregate and light-harvesting antenna take the value of $\sim 50$cm$^{-1}$ which is $\sim 0.3\%$ of the exciton frequency. This is much stronger nonlinearity than that in YIGs with Kerr coefficient $K\sim 0.1$nHz that is $\sim 10^{-11}$ of its Kittel frequency. Recent development in both ultrafast spectroscopy and synthesis have revealed the important role of quantum coherence which may significantly modify the functions of complex molecules and may help the design of polaritonic molecular devices as well as polariton chemistry. Hence entangling the molecular aggregates may help the studies of quantum phenomena in complex molecules.

We gratefully acknowledge the support of AFOSR Award FA-9550-18-1-0141, ONR Award N00014-16-1-3054 and Robert A. Welch Foundation (Award A-1261 \& A-1943-20180324). We also thank Jie Li and Tao Peng for the useful discussions.

\end{document}